\documentstyle[12pt]{article}
\begin{document}
\setlength{\topmargin}{-1cm}
\setlength{\oddsidemargin}{0cm}
\setlength{\evensidemargin}{0cm}
\title{
\begin{flushright}
{\small CERN-TH/96-286}
\end{flushright}
\vspace{1cm}
 {\bf The Role of the Anomalous $U(1)_A$ for the Solution of the
Doublet--Triplet Splitting Problem}}
 
\author{{\bf Gia Dvali}\thanks{E-mail:
georgi.dvali@cern.ch}\\ CERN, CH-1211 Geneva 23, Switzerland\\
\and
{\bf Stefan Pokorski}\\Institute
for Theoretical Physics, Warsaw 
University, Poland\\and\\ Max-Planck-Institut f\"{u}r Physik,
F\"{o}hringer Ring 6'\\ 80805 Munich, Germany\\}
 
\date{ }
\maketitle
 
\begin{abstract}
The anomalous $U(1)_A$ symmetry provides a generic method of
getting accidental symmetries. Therefore, it can play a crucial role
in solving the doublet--triplet splitting and the $\mu$-problems via
the {\it pseudo-Goldstone} mechanism: $U(1)_A$
can naturally uncorrelate the two
grand unified Higgs sectors in the superpotential
and simultaneously induce the desired
expectation values through the Fayet--Iliopoulos $D$-term. The zero
modes of the resulting compact vacuum 
degeneracy can be identified with
the massless electroweak Higgs doublets. This automatically solves
the doublet--triplet splitting and the $\mu$-problems
to all orders in $M_P^{-1}$.
No additional discrete or global symmetries are needed. $U(1)_A$ can
also play the role of the `matter parity' and suppress the
baryon number violating
operators. We present the simplest $SU(6)$ gauge model with a minimal
Higgs sector and no doublet--triplet splitting problem. This model also
relates the fermion mass hierarchy to the hierarchy of scales and
predicts, for the generic non-minimal K\"ahler potential,
approximately universal values for the tree-level
electroweak Higgs mass parameters.

\end{abstract}
 
\newpage
 
\subsection*{1. Introduction}

One of the most difficult problems of the supersymmetric grand unified
theories (GUTs) is the doublet--triplet splitting problem. It is
difficult to understand
how the theory, which knows only the very large scales 
$M_G \sim 10^{16}$ GeV and $M_P \sim 10^{19}$ GeV,
arranges itself in such a way
that pair of essentially massless electroweak doublets $H, \bar H$
survive down to the low energies, not accompanied by their colour triplet
partners. The natural logic is to 
attribute the lightness of the Higgs doublets
to the smallness of the supersymmetry-breaking scale in the low energy
sector $m_{3/2} \sim 100$ GeV. This requires a mechanism that would
ensure masslessness of the doublets in the supersymmetric limit and
at the same time guarantee that desired mass terms ($\mu$ and
$B\mu$) of the right order of magnitude are generated
by the SUSY breaking.
As a guideline we will follow this strong criterion of naturalness,
according to which the single mechanism must be responsible for
both: {\it i}) vanishing doublet mass in the SUSY limit and
{\it ii}) appearance of $\mu^2
\sim B\mu \sim m^2_{3/2}$ after its breaking.
We also adopt the $minimality$ requirement:
both problems must be solved
within the minimal set of the
Higgs fields needed to break the GUT symmetry
to the standard model. Besides the aesthetic problems, the non-minimal
Higgs sector (additional adjoints etc.) usually creates difficulties with 
asymptotic freedom.
As far as we know, the only approach that can satisfy the
above criterion is the `pseudo-Goldstone'
idea \cite{u5, u61, u62, bdsbh, u63}.
The key point is that Higgs doublets can be identified
as the zero modes of the compact vacuum degeneracy, which
are massless to all orders in perturbation theory,
because of supersymmetry.
Once supersymmetry is broken, the flat directions are lifted and
the doublets get masses  of just the
right order of magnitude: $\sim m_{3/2}$.
On the way to constructing a realistic model
along these, lines there are a
few potential difficulties: 1) flat direction should not be a result
of the fine tuning, but rather be guaranteed by the exact
symmetries of the theory; 2) unless it is protected by the gauge
symmetries, the flat direction can be lifted by the $M_P$-suppressed
operators in the superpotential, which can destroy the original solution;
3) colour-triplet partners must be heavy and decouple along
the flat direction.

Closer to the realization of this
program came the model of\cite{u61, u62}.
The crucial observation was that
the desired compact degeneracy, automatically
satisfying condition (3) above, could result if the different Higgs
fields that break GUT symmetry are
not correlated (have no cross-couplings)
in the superpotential. In this case, the
vacuum has an accidental flat direction
corresponding to the independent global rotations of the uncorrelated
vacuum expectation values (VEVs).
Since this rotation is not an exact symmetry
of the theory (it is broken by the gauge and Yukawa couplings)
the corresponding zero modes are not eaten up by any gauge field and
are physical.

 Thus, the central issue is to suppress the unwanted cross-couplings by 
exact symmetries. Here one can identify
the following problems: first, the
symmetries, which forbid the
cross-couplings, also restrict the possible
self-couplings of one of the fields, so that its VEV vanishes
and the flat direction disappears;
secondly these symmetries are anomalous
and cannot be ordinary gauge symmetries. Thus, there is no reason why
they should be respected by the Planck scale suppressed, operators which
would generate an unacceptably large mass for the doublets.

 The key point of the present paper is
that the anomalous gauge $U(1)_A$
symmetry, usually present in string theories \cite{Dterm}, can provide
a simultaneous solution to the above
problems. Cancellation of the anomalies
by the Green--Schwarz mechanism \cite{gs}
requires non-zero mixed anomalies
and thus, some of the GUT fields must transform 
under $U(1)_A$. Since the symmetry is anomalous,
the Fayet--Iliopoulos term
(proportional to the sum of charges ${\rm Tr} Q$)
is always generated \cite{witten} and, in strings,
is given by\cite{Dterm}
\begin{equation}
\xi = {g^2 {\rm Tr}Q \over 192\pi^2}M_P^2.
\end{equation}

Since it is a gauge symmetry, the anomalous $U(1)_A$ can naturally 
uncorrelate the GUT VEVs in the superpotential to {\it all orders}
in $M_P^{-1}$ and
at the same time induce the desired VEV $\sim \sqrt{\xi}$
through the
Fayet--Iliopoulos $D$-term. This gives an exciting possibility of
solving the doublet--triplet
splitting and the $\mu$-problems in all orders in
$M_P^{-1}$, without any need of
additional discrete or global symmetries,
and within the minimal Higgs content. Incidentally it turns out
that in this approach $U(1)_A$ plays the role of the matter parity
also and can suppress all the dangerous baryon number violating operators.

Previously the implications of the
anomalous $U(1)_A$ were considered for the
fermion \cite{fermion} and sfermion \cite{sfermion} masses, for
mediating the supersymmetry breaking, and for the  flavour
problem \cite{dpbd}. Here we show that it is a new and crucial role
that $U(1)_A$ can play for the solution of the doublet--triplet
splitting and the
$\mu$-problems.\footnote{The possibility of suppressing
the $\mu$-term in the superpotential via the anomalous $U(1)_A$ and
generating it through the couplings in the K\"ahler potential, via the
Giudice--Masiero mechanism \cite{gm}, has been considered \cite{othermu}.
This approach, however, requires an additional mechanism for
the doublet-triplet splitting problem and will not be considered here}  

\subsection*{2 The Problem and the Solution}

To illustrate the problem and our solution we will consider the model
of ref.\cite{u61, u62}. Consider the minimal supersymmetric $SU(6)$ GUT.
In order to break the symmetry down to the standard model group
$G_W = SU(3)_C\otimes SU(2)_L\otimes U(1)_Y$, a minimum of two Higgs
representations are necessary: an adjoint $\Sigma_i^k$ and a
fundamental--antifundamental pair $\bar{\Phi^i},\Phi_i$
($i,k = 1,2,...,6$). The relevant $D$-flat VEVs are:
\begin{equation}
\Sigma = {\rm diag}(1,1,1,1,-2,-2)\sigma,~~~
\Phi_i = \bar{\Phi^i} = (\phi,0,0,0,0,0),
\end{equation}
which leave unbroken $G_{\Sigma} = SU(4)_c\otimes SU(2)_L\otimes U(1)$
and $G_{\Phi} = SU(5)$ symmetries respectively, so that the intersection
gives unbroken $G_W$. Assume now that these two sectors have no
cross-couplings in the superpotential
\begin{equation}
W = W(\Sigma) + W(\Phi).
\end{equation}
Thus, it effectively has $G_{gl} = SU(6)_{\Sigma}\otimes SU(6)_{\Phi}$
symmetry. Since for the VEVs given in Eq(2) this global symmetry
is broken to $G_{\Sigma}\otimes G_{\Phi}$,
 there are compact flat directions
in the vacuum that do not correspond to any broken gauge generator;
thus, the corresponding zero modes are  physical fields.
Note that the $SU(6)$ $D$-terms cannot lift this degeneracy, since the
contributions of $\Phi,\bar{\Phi}$ and $\Sigma$ are {\it independently}
zero.
By a simple counting of the Goldstone states and
of the broken gauge generators,
we find that left-over zero modes are two linear combinations of
the electroweak doublets from $\Sigma$ and $\Phi$ ($\bar {\Phi}$):
\begin{equation}
H = {H_{\Sigma}\langle \phi \rangle - H_{\Phi}3\langle\sigma \rangle 
\over \sqrt{\langle \phi \rangle^2 + 9\langle \sigma \rangle^2}},~~~~~
\bar H = {\bar H_{\Sigma}\langle \phi \rangle - \bar H_{\Phi}3
\langle\sigma \rangle \over \sqrt{\langle \phi \rangle^2 + 
9\langle \sigma \rangle^2}}.
\end{equation}
All other states are heavy and the doublet--triplet splitting problem is
solved. The main difficulty is to justify the absence of the possible
cross-couplings in the superpotential up to a sufficiently high order in
$M_P^{-1}$,
by some exact symmetry. This is very difficult to do without also
forbidding the possible self-couplings of the Higgs fields, so
that usually one ends up either with one of the VEVs being zero, or
with an enormous degeneracy of the vacuum, with many new, massless,
coloured and
charged superfields. More importantly, perhaps, the global
symmetries under which the cross-coupling $\Sigma\bar{\phi}\phi$ is
non-invariant are anomalous and need not be respected by $M_P^{-1}$
suppressed operators. Any such mixed operator with dimensionality 
less than 6 -- 7 would destroy the solution completely. (Note,
the higher operators are safe only if $\phi << M_P$, which is
an additional input of the theory.) This consideration indicates
that we are naturally lead,
in the problem of separating the two sectors,
to the concept of  anomalous gauge symmetry. As we
now show, the
$U(1)_A$ symmetry provides a natural loophole due to the simple
reason that it is `anomalous'. It is enough to assume that
$\Phi$,$\bar{\Phi}$ fields carry negative charges $q$ and $\bar q$
and all the other fields, and in particular quarks and leptons,
carry non-negative charges so that the total trace
${\rm Tr Q} > 0$.
As we will see below, this assumption naturally fits in the structure of
Yukawa couplings and also avoids dangerous charge and colour breaking
flat directions. We also assume that $\Sigma$ carries zero charge.
Then $\Phi$ and $\bar{\Phi}$ are simply left out of the most general
$SU(6)\otimes U(1)_A$-invariant Higgs superpotential\footnote{Since
our discussion goes at the level of an effective field theory below
$M_P$, we have simply presented the most general superpotential. Its
precise form is not important for our purposes provided it can induce the
desired VEV $\sigma \sim M_G$. This is just a necessary condition
for having the GUT below the Planck scale. In reality we
assume the scale $M \sim M_G$ to be dynamically generated from $M_P$
by some yet unexplored mechanism}
\begin{equation}
W_{Higgs} = {M\over 2}\Sigma^2 + {h\over 3}\Sigma^3 +
\lambda_n{\Sigma^n \over M_P^{n - 3}},
\end{equation}
which fixes the VEV as in Eq(1) with 
$\sigma = {M \over h}\left( 1 + O(M_G/M_P) \right )$.
The VEV of the $\phi$ is fixed from the $D$-terms
\begin{equation}
{g^2 \over 2}\left [\Phi^*T^a\Phi - \bar{\Phi}T^a\bar{\Phi}^* +
[\Sigma^*\Sigma]T^a + {\rm matter ~ fields}\right ]^2 + 
{g^2_A \over 2}
\left [q|\Phi|^2 + \bar q|\bar{\Phi}|^2  + \xi + q_i|S_i|^2\right ]^2,
\end{equation}
where $T^a$ are $SU(6)$-generators and $q_i|S_i|^2$ is a sum  over all the
positively charged fields with $q_i > 0$. Minimization gives
$\phi^2 = -\xi /(q + \bar q)$ (the equality $\Phi = \bar{\Phi}$ is demanded
from the $SU(6)$ $D$-term).
The only allowed cross-couplings between $\Sigma$ and $\Phi$ sectors are
the ones that involve positively charged matter field (see Yukawa
couplings below). These couplings, however, can never affect the
vacuum degeneracy, since all the positively charged fields have
{\it zero} VEVs.
Thus, the doublet--triplet splitting problem is solved in
all order in $M_P^{-1}$ without need of any extra symmetries.

\subsection*{ 3 $\mu$ and $B\mu$}

Assuming the conventional \cite{gravity} gravity-mediated hidden
sector supersymmetry breaking, both $B\mu \sim \mu^2$ of the
desired magnitude are automatically generated in this scenario and
we end up with
the following tree-level relation among the electroweak Higgs
doublet mass parameters
\begin{equation}
m_{H}^2 \simeq m_{\bar H}^2 \simeq B\mu = \mu^2 + m^2, ~~~
\mu = (3A_{(3)}/h - 2A_{(2)}),
\end{equation}
where $m$ is a soft
mass of the $\Sigma$ field and $A_{(3)}$ and $A_{(2)}$ are
coefficients of the
soft trilinear and bilinear couplings, respectively.
The above relation is given at $M_P$ and holds up to the corrections of
order $\epsilon = {\xi \over M_P^2}$.
This is a standard pseudo-Goldstone relation of \cite{u5,u61,u62} for the
minimal soft terms and is due to the fact that for the minimal K\"ahler
potential at the tree level there
should be one exactly massless state
\begin{equation}
H_+ = { H + \bar H^* \over \sqrt {2}}
\end{equation}
In our case this relation
holds for the arbitrary non-minimal K\"ahler potential and essentially
is a prediction of the model. This is
because in our model the scale $\phi \sim \sqrt{\xi}$ is predicted to be
just half-way between
$M_P$ and $M_G$ and the light pseudo-Goldstones predominantly
reside in $\Sigma$. 
Due to this, both  contributions from the non-universal
soft terms of $\Phi$ and $\bar{\Phi}$ and contributions from the
cross couplings in the K\"ahler potential are suppressed.
Thus, we have in this model one less free parameter than in minimal
supergravity; hence it can predict, for instance, tan$\beta$
in terms of the other masses\cite{u5, u62, bolo}

\subsection*{4 The Fermion Masses and Proton Stability}

The fermion masses in the above scheme were analysed in more detail in
\cite{bdsbh}, where it was shown that the model admits a realistic
(within uncertainties in coupling constants of order 1) description
of the fermion mass hierarchy in terms of the hierarchy of scales
$M_P >> \phi >> M_G$ without invoking flavour symmetries.
The most interesting result is
that only the top quark has a renormalizable
Yukawa interaction at the tree level. This happens if besides the three
chiral families in $15_{\alpha} + \bar 6'_{\alpha} + \bar 6_{\alpha}$
(the minimal anomaly-free set that accommodates
$10 + \bar 5$ of $SU(5)$ per family)\footnote{$15 + \bar 6 +\bar 6$ just
compose a fundamental $27$-plet of $E_6$, in the case of the embedding
of $SU(6)$ into $E_6$.}
one assumes an odd number of real $20$-plets
with invariant $M_P$ mass terms.
A decomposition of these multiplets in terms of $SU(5)$ representations
gives
\begin{equation}
15_{\alpha} = 10_{\alpha} + 5_{\alpha},~~ 20 = 10 + \bar{10}
\end{equation}
The important group-theoretical fact is that no invariant mass term
can be formed from the symmetric product of two $20$-plets; thus,
a single $20$-plet will survive as light
and can get a mass only after $SU(6)$ symmetry breaking.
Up to a field redefinition, the most general renormalizable couplings
are (coupling constants are neglected)
\begin{equation}
\Sigma 20 20 + \Phi 15_3 20 + \bar{\Phi}15_{\alpha}\bar 6_{\beta}'.
\end{equation}
The last coupling simply gives $SU(5)$-invariant masses to the
extra heavy states ($5,\bar 5'$)
from $15$-s and $\bar 6'$-s, mixing them with each other. The second term
combines a $10_3$-plet from $15_3$ with $\bar{10}$ from
$20$, and they
become heavy as well. The remaining light $10$, predominantly residing in
$20$, gets a tree-level
Yukawa coupling with $H$ through the first term, giving mass to the
top. The masses of lighter fermions are generated through
the higher-dimensional operators
\begin{equation}
{1 \over M_P^{n + 1}}\Phi\Sigma^n\Phi 15 15 +
{1 \over M_P^n}\bar{\Phi}\Sigma^n 15\bar{6} + ...
\end{equation}
with different possible $\Sigma$ insertions.
This gives us the possibility to
account for the fermion mass
hierarchy in terms of two ratios of the scales
${M_G \over M_P}$ and ${\phi \over M_P}$; for more details we refer the 
reader to
\cite{bdsbh}. The only new point
in our case is that the necessary condition
$M_G << \phi << M_P$, which was an
additional input of the theory in the
case of \cite{bdsbh}, is now a natural outcome  since the scale
$\phi$ is generated from the Fayet--Iliopoulos $D$-term.
It is easy to show that our $U(1)_A$ charge assignment (necessary to
solve the doublet--triplet
splitting problem) is automatically compatible
with the above structure of
Yukawa couplings. The simplest possibility is not
to invoke any flavour dependence
in the spirit of ref \cite{bdsbh}. Then the
flavour-blind $U(1)_A$ charges are constrained as
\begin{equation}
q_{15} = -q, ~~ q_{\bar 6} = q - \bar{q}.
\end{equation}
The additional constraint comes
from the neutrino masses. For example, if
we generate the  right-handed  neutrino masses from the operator
\begin{equation}
{(\Phi \bar 6)^2 \over M_P^3}(\Phi \bar {\Phi}),
\end{equation}
then charges are fixed as
$q_{15} = -q, ~~ q_{\bar 6} = -4q, ~~  \bar{q} = 5q$.
This assignment automatically kills any baryon number violating operator
trilinear in the
matter fields $\bar 6 15 \bar 6$ to all orders in $M_P^{-1}$.
Thus, $U(1)_A$ can play the role of the matter parity.
Family-dependent charge assignment, along the lines of \cite{fermion},
is also possible without altering any of our conclusions.
The novel feature in such a construction, not attempted here, will be
that in contrast to \cite{fermion} the Higgses that break $U(1)_A$ are
not the GUT singlets. 
hus their Yukawa couplings will be constrained both, by
the GUT symmetry and by the anomalous $U(1)_A$. This can offer the the 
possibility of
generating specific (and hopefully predictive) 
textures for the fermion masses.

\subsection*{Acknowledgements}
We thank Alex Pomarol for discussions.
S.P. was partially supported by the Polish Committee for Scientific
Research.


\begin{thebibliography}{999999999}



\bibitem{u5} K.Inoue, A.Kakuto and H.Takano,
{\it Prog. Theor. Phys.} {\bf 75} (1986) 664;
A.Anselm and A.Johansen, {\it Phys. Lett.}{\bf B200} (1988), 331;
A.Anselm, {\it Sov. Phys. JETP} {\bf 67} (1988), 663.
In the $SO(10)$ framework see, R.Barbieri, G.Dvali and A.Strumia,
{\it Nucl. Phys.} {\bf B391} (1993) 487.

\bibitem{u61} Z.Berezhiani and G.Dvali, {\it Sov. Phys. Lebedev Inst.
Rep.} {\bf 5} (1989) 55.

\bibitem{u62} R.Barbieri, G.Dvali and M.Moretti, {\it Phys. Lett.}
{\bf B312} (1993) 137.


\bibitem{bdsbh} R.Barbieri, G.Dvali, A.Strumia, Z.Berezhiani and
L.Hall, {\it Nucl. Phys.} {\bf B432} (1994) 49.

\bibitem{u63} Z.Berezhiani, C.Cs\'aki and L.Randall, {\it Nucl.Phys.}
{\bf B444} (1995) 61; Z.Berezhiani, {\it Phys. Lett.}{\bf B355}
(1995) 481; C.Cs\'aki and L.Randall, hep-ph/9512278;
B.Ananthanarayan and Q.Shafi, hep-ph/9512345.

\bibitem{Dterm} M.Dine, N.Seiberg, and E.Witten, {\it Nucl. Phys.}
{\bf B289} (1987) 585; J.Atick, L.Dixon, and A.Sen,
{\it Nucl. Phys.} {\bf B292} (1987) 109;
M.Dine, I.Ichinose, and N.Seiberg,
{\it Nucl. Phys.} {\bf B293} (1987) 253.


\bibitem{gs} M.Green and J.Schwarz,
{\it Phys.Lett.} {\bf B149} (1984) 117.

\bibitem{witten} E.Witten,
{\it Nucl. Phys.} {\bf B 188}(1981) 513; W.Fischler
H.P.Nilles, J.Polchinski, S.Raby, L.Susskind,
{\it Phys. Rev. Lett.} {\bf 47} (1981) 757.


\bibitem{fermion} See, for example, L.Ib\'anez and G.G. Ross,
{\it Phys. Lett.} {\bf B332} (1994) 100;
V.Jain and R.Shrock, hep-ph/9412367,
{\it Phys.Lett.} {\bf B352} (1995) 83;
P.Bin\'etruy and P.Ramond,
{\it Phys. Lett.} {\bf B350} (1995) 49;

E.Dudas, S.Pokorski, and C.A.Savoy,
{\it Phys. Lett.} {\bf B356} (1995) 45;
for the recent analysis of both fermion and sfermion masses see,
E.Dudas {\it et al} in \cite{sfermion}.

\bibitem{sfermion} H.Nakano, hep-th/9404033; 
E.Dudas, S.Pokorski, and C.A.Savoy,
{\it Phys.Lett.} {\bf B369} (1996) 255;
E.Dudas, C.Crojean, S.Pokorski, and C.A. Savoy, hep-ph/9606383;
Y.Kawamura and T.Kobayashi, {\it Phys.Lett.} {\bf B375} (1996) 141.

\bibitem{dpbd} G.Dvali and A.Pomarol, Preprint CERN-TH/96-192,
hep-ph/9607383, to appear in {\it Phys. Rev. Lett.};
P.Bin\'etruy and E.Dudas, hep-th/9607172.

\bibitem{gm} G.F.Giudice and A.Masiero, {\it Phys.Lett.} {\bf B206}
(1988) 480.

\bibitem{othermu} Y.Nir, {\it Phys. Lett.} {\bf B354} (1995) 107;
V.Jain and R.Shrock, hep-ph/9507238;
T.Gherghetta, G.Jungman and E.Poppitz, hep-ph/9511317.
For recent attempts see, e.g.,
A.G.Cohen, D.B.Kaplan and A.E.Nelson, hep-ph/9607394; R.N.Mohapatra
and A.Riotto, hep-ph/9608441. 

\bibitem{gravity} R.Barbieri, S.Ferrara and C.A.Savoy, {\it Phys.Lett.}
{\bf B119} (1982) 343; A.H. Chamseddine, R.Arnowitt and P.Nath,
{\it Phys.Rev.Lett} {\bf B49} (1982) 970; H.-P.Nilles, M.Srednicki and
D.Wyler, {\it Phys.Lett.} {\bf B120} (1983) 346; L.Hall, J.Lykken and
S.Weinberg, {\it Phys.Rev.} {\bf D 27} (1983) 2359.

\bibitem{bolo} For the recent analysis see last two references in \cite{u63}.
\end{thebibliography}
\end{document}